\documentclass{ptephy_v1}%%%%%% to generate preprint number with ptep logo

\usepackage{amssymb}
\usepackage{amsmath,amscd}
\usepackage{color}
\usepackage{mathrsfs}
\usepackage{mathtools}

\usepackage{bm}
\usepackage[normalem]{ulem}
\newcommand{\D}{{\cal D}}

\begin{document}

\title{Divergence equations and uniqueness theorem of static spacetimes with conformal scalar hair}

%%%% To generate auto affiliation numbers please use \author{}\affil{} command

%<<<<<<<<<<<<< AUTHOR  >>>>>>>>>>>>>>>%
\author{Takeshi Shinohara${}^1$}
\author{Yoshimune Tomikawa${}^2$}
\author{Keisuke Izumi${}^{3,1}$}
\author{Tetsuya Shiromizu${}^{1,3}$}

%<<<<<<<<<<<<< AFFILIATION >>>>>>>>>>>>>>>%
\affil{${}^1$Department of Mathematics, Nagoya University, Nagoya 464-8602, Japan}
\affil{${}^2$Faculty of Economics, Matsuyama University, Matsuyama 790-8578, Japan}
\affil{${}^3$Kobayashi-Maskawa Institute, Nagoya University, Nagoya 464-8602, Japan}

%%% To include the collaborator name... Please use the command "\collaborator"
%%% For example: \collaborator{ATLAS Collaboration}

\begin{abstract}
We reexamine the Israel-type proof of the uniqueness theorem of the static spacetime outside the 
photon surface in the Einstein-conformal scalar system. 
We derive in a systematic fashion a new divergence identity which plays a key role in the proof. 
Our divergence identity includes three parameters, allowing us to give a new proof of the uniqueness. 
\end{abstract}

\subjectindex{E0}

\maketitle

%======================================%
%<<<<<<<<<<<< SECTION I  >>>>>>>>>>>>>>%
%======================================%
%
\section{Introduction}

The Bocharova-Bronnikov-Melnikov-Bekenstein (BBMB) solution \cite{BBM, Bekenstein} is a static black hole 
solution to the Einstein-conformal scalar system in four dimensions.\footnote{Assuming the analyticity  
at the photon surface, it was shown that higher dimensional counterpart fails to admit a regular event horizon \cite{HDBBMB}.} 
A natural question to be asked is whether 
this solution exhausts all the static black holes in this theory. The original uniqueness proof  \cite{Israel,Robinson} 
of static black holes in vacuum general relativity demonstrates the uniqueness of the boundary value problem of 
elliptic system between the event horizon and spatial infinity. In the BBMB solution, the photon surface composed of 
(unstable) closed circular orbits of photons appears at the points where the coefficient of the Ricci tensor vanishes
in the Einstein equation. This feature prevents us to apply the global boundary value problem outside the event horizon. 
Nevertheless, the uniqueness property of the  static region outside the photon surface  has been properly addressed in 
Refs. \cite{Tomikawa2017a,Tomikawa2017b}, where it has been shown to be isometric to the BBMB solution. 

To prove the uniqueness theorem, two technically and conceptually distinct methods are available so far. 
The BBMB uniqueness has been demonstrated 
in Ref. \cite{Tomikawa2017a} by a way similar to those in Refs. \cite{Israel,Robinson}, 
relying on certain divergence identities. 
The other proof in Ref. \cite{Tomikawa2017b} follows the argument 
in Ref. \cite{bm} based on the conformal transformation and positive mass theorem \cite{SY}. 
Meanwhile, 
for the uniqueness of black holes in vacuum Einstein, Einstein-Maxwell and Einstein-Maxwell-dilaton systems, 
the argument by Robinson \cite{Robinson}, which is regarded as a simplification of Israel's proof, has been reexamined in Ref. \cite{Nozawa2018}. A significant achievement in Ref. \cite{Nozawa2018} is to provide 
a systematic way to derive the divergence identities exploiting the proper deviation from the Schwarzschild metric. 
The obstruction tensors are of great use in finding a series of divergence identities even in stationary metrics \cite{Nozawa:2021udg}.
Then, it is natural to ask if the procedure developed in Ref. \cite{Nozawa2018} 
works  in the Einstein-conformal scalar system and also for the uniqueness proof of photon surfaces. 

In this paper, we apply the procedure of Ref. \cite{Nozawa2018} to the Einstein-conformal scalar system. 
We shall see that it indeed works and find a new divergence identity with three parameters. 
Since the derivation for the divergence 
identities found in Ref. \cite{Tomikawa2017a} was rather non-trivial, the  systematic way to derive the identity will be of some help in a similar consideration for 
other systems. Finally, we shall prove the uniqueness of the static photon surface again. 

The rest of this paper is organized as follows. In Sect. 2, we describe the Einstein-conformal scalar system and 
the setup of the current paper. In Sect. 3, we develop the procedure of Ref. \cite{Nozawa2018} to the 
Einstein-conformal scalar system. Finally, we will give summary in Sect. 4. In Appendix, we present the relation 
to Ref. \cite{Tomikawa2017a} in the details. 

%
%======================================%
%<<<<<<<<<<<< SECTION II  >>>>>>>>>>>>>%
%======================================%
\section{The BBMB black hole and setup}

In this section, we describe the Einstein-conformal scalar system and the basic equations for static spacetimes.
The action for the Einstein-conformal scalar system is represented by 
\begin{align}
\label{actionconf}
S=\int d ^4 x \sqrt{-g}\left(\frac 1{2\kappa}
R -\frac 12 (\nabla \phi)^2 -\frac 1{12} R 
\phi^2 \right),
\end{align}
where $\kappa=8\pi G$ is a gravitational constant. 
The field equations are given by 
\begin{eqnarray}
\left(1-\frac \kappa 6 \phi^2\right)\left(R_{\mu\nu}
-\frac12 R g_{\mu\nu}\right)=&\kappa \left( \nabla_\mu \phi \nabla_\nu \phi-\dfrac12 g_{\mu\nu}(\nabla \phi)^2
+\dfrac16 (g_{\mu\nu} \nabla^2-\nabla_\mu \nabla_\nu)\phi^2 \right) \label{Eineq0}
\end{eqnarray}
and 
\begin{equation}
\nabla^2 \phi-\frac{1}{6}R\phi=0\label{scalar}.
\end{equation}
Taking the trace of Eq. (\ref{Eineq0}), one finds $R=0$, so that the 
field equations are simplified to 
\begin{equation}
\left(1-\frac{\kappa}{6}\phi^2\right)R_{\mu\nu}=\kappa \left( \frac{2}{3}\nabla_{\mu}\phi\nabla_{\nu}\phi-\frac{1}{6}g_{\mu\nu}
(\nabla\phi)^2-\frac{1}{3}\phi \nabla_{\mu}\nabla_{\nu}\phi \right)
\label{einstein}
\end{equation}
and 
\begin{equation}
\nabla^2 \phi=0 .\label{scalar2}
\end{equation}

This theory admits a static black hole solution (the BBMB solution)
\cite{BBM,Bekenstein}. Its metric and scalar field are 
\begin{equation}
ds^2=-\left(1-\frac{m}{r}\right)^2dt^2 + \left(1-\frac{m}{r}\right)^{-2}dr^2 + r^2 d\Omega^{2}_{2}\label{bbmb}
\end{equation}
and
\begin{equation}
\phi=\pm {\sqrt {\frac{6}{\kappa}}}\frac{m}{r-m},
\end{equation}
where $d\Omega^2_2$ is the metric of the unit two-sphere and $m$ is the mass which is supposed to 
be positive. The event horizon is located at $r=m$. 

One important feature of the Einstein-conformal scalar system is that it may admit points satisfying 
$\phi=\pm \sqrt{6/\kappa}$, where the prefactor of the Ricci tensor in (\ref{einstein}) vanishes. 
This means that the effective gravitational constant diverges.  
For the BBMB solution, this occurs precisely at the photon surface $r=2m$. As far as the outside 
region of the photon surface is concerned,  the uniqueness property has been settled to be affirmative 
\cite{Tomikawa2017a,Tomikawa2017b}. As stated in Sect. 1, 
we will reexamine the proof of Ref. \cite{Tomikawa2017a} and then present an elegant way to find the 
divergence identities used in the proof. 

The generic form of a static metric is written as 
\begin{equation}
ds^2=-V^2(x^{k})dt^2 + g_{ij}(x^{k})dx^i dx^j \label{static},
\end{equation}
where $V$ is the norm of the static Killing vector. 
The event horizon located at $V=0$. We also assume that the conformal scalar field is also static $\phi=\phi(x^i)$.
The Einstein equation becomes 
\begin{equation}
\Bigl(1-\frac{\kappa}{6}\phi^2 \Bigr)VD^2 V= \frac{\kappa}{6}\left[V^2 (D\phi)^2+2\phi V D^i V D_i \phi\right] \label{einstein00}
\end{equation}
and
\begin{eqnarray}
\Bigl(1-\frac{\kappa}{6}\phi^2 \Bigr)\Bigl({}^{(3)}R_{ij}-V^{-1}D_i D_j V\Bigr)
=\kappa\left(\frac{2}{3}D_i \phi D_j \phi -\frac{1}{6}g_{ij}(D\phi)^2-\frac{1}{3}\phi D_i D_j \phi
\right), 
 \label{einsteinij}
\end{eqnarray}
where $D_i$ and ${}^{(3)}R_{ij}$ are  the covariant derivative and  the Ricci tensor with respect to 
the three dimensional metric $g_{ij}$, respectively. Note here that the front factors in the left-hand side 
of Eqs. (\ref{einstein00}) and (\ref{einsteinij}) vanish at the surface $S_p$ determined by 
$\phi=\pm \sqrt{6/\kappa}=:\phi_p$. 
The equation for the scalar field is written as 
\begin{equation}
D_i (VD^i \phi)=0 \label{staticscalar}.
\end{equation}
The asymptotic conditions at infinity are given as
\begin{equation}
V=1-\frac{m}{r}+\mathcal{O}\left(\frac{1}{r^2}\right),\label{asympV}
\end{equation}
\begin{equation}
g_{ij}=\left(1+\frac{2m}{r}\right)\delta_{ij}+\mathcal{O}\left(\frac{1}{r^2}\right) \label{asympg}
\end{equation}
and
\begin{eqnarray}
\phi=\mathcal{O}\left(\frac{1}{r}\right).\label{asympphi}
\end{eqnarray}

Equations (\ref{einstein00}) and (\ref{staticscalar}) give us  
\begin{equation}
D_i[(1-\varphi)D^i \Phi]=0,\label{DDPhi}
\end{equation}
where $\Phi:=(1+\varphi)V$ and $\varphi=\pm {\sqrt {\kappa/6}}\phi$. 
Then, one considers $\Omega$ which is the bounded region by $S_p$ and the two-sphere $S_\infty$ at 
spatial infinity. The volume integration of Eq. (\ref{DDPhi}) over $\Sigma$ shows 
the relation between 
$V$ and $\phi$ as \cite{Tomikawa2017a} 
\begin{equation}
\phi=\pm \sqrt{\frac{6}{\kappa}}(V^{-1}-1).\label{relation}
\end{equation}
Through Eq. (\ref{relation}), we see that $V=1/2$ at $S_p$

Using the relation (\ref{relation}), the Einstein equation implies 
\begin{equation}
D^2 v=0, \label{lapsev}
\end{equation}
where $v \coloneqq \ln V$. Henceforth, we can regard $v$ as a kind of the radial coordinate. 
It allows us to decompose the $t=$constant hypersurface $\Sigma$ into the radial direction and the foliation of 
the $v=$ constant surfaces $S_v$.
As a consequence, 
the $(i,j)$-component of the Einstein equation becomes
\begin{equation}
(2V-1){}^{(3)}R_{ij}=D_i D_j v + (4V+1) D_i v D_j v -g_{ij} (Dv)^2.\label{einsteinij2}
\end{equation}
%And we can take the foliation of the $v=$ constant surfaces $S_v$ on the $t=$constant hypersurface $\Sigma$. 
The curvature invariant is expressed in terms of geometrical quantities associated with $S_v$ as 
\begin{eqnarray}
R_{\mu\nu}R^{\mu\nu}
& = & \frac{1}{(2V-1)^2 \rho^2}\left[\left(2(1-V)k_{ij}-\frac{1}{\rho}h_{ij}\right)^2
+\left(-2(1-V)k+\frac{1+2V}{\rho}\right)^2\right. \nonumber\\
& &~~+\left. \frac{8(1-V)^2}{\rho^2}(\D \rho)^2\right]+\frac{1}{\rho^4},\label{ricciinv}
\end{eqnarray}
where $h_{ij}$ is the induced metric of $S_v$, ${\cal D}_i$ is the covariant derivatve with respect to $h_{ij}$ 
and $\rho$ is the lapse function, $\rho:=(D_ivD^iv)^{-1/2}$. 
Moreover, $k_{ij}$ is the extrinsic curvature of $S_v$ defined by $k_{ij}:=h_i^kD_kn_j$, where $n_i$ is the 
unit normal vector to $S_v$ on $\Sigma$, and $k$ is the trace part of $k_{ij}$. Using the lapse function, $n_i$ is 
expressed by $n_i=\rho D_i v$. From Eq. (\ref{ricciinv}), at $V=1/2$, we have to impose 
\begin{equation}
\D_i \rho |_{S_p}=0,\ \ \ \ k_{ij}|_{S_p}=\frac{1}{\rho_p}h_{ij}|_{S_p},\label{BCatSp}
\end{equation}
otherwise the curvature invariant diverges. 
The first equation of Eq. (\ref{BCatSp}) shows that $\rho$ is constant on $S_p$. 
We write the constant as $\rho_{p}\coloneqq \rho|_{S_p}$. 
The second condition of Eq. (\ref{BCatSp})  implies that the surface $S_p$ is totally umbilic. 
We can see that, under the conditions of Eq. (\ref{BCatSp}),  the Kretschmann invariant, $R_{\mu\nu\rho\sigma}R^{\mu\nu\rho\sigma}$ 
is also finite at $S_p$ \cite{Tomikawa2017a}. 

In the proof of the uniqueness in Ref. \cite{Tomikawa2017a} the following divergence identities are presented without any explanations, 
\begin{equation}
D_{i}\left(\frac{n^i}{\rho}\right)=0,\label{tmkw1}
\end{equation}
\begin{equation}
D_i \left[\frac{(\rho k -2)n^i}{(2V-1)\rho^{\frac{3}{2}}}\right]=-\frac{1}{2V-1}
\rho^{-\frac{1}{2}}(\Tilde{k}_{ij}\Tilde{k}^{ij}+\rho^{-1}\D^2 \rho)\label{tmkw2}
\end{equation}
and
\begin{equation}
D_i ((k \xi + \eta)n^i)=-(\Tilde{k}_{ij}\Tilde{k}^{ij}+\rho^{-1}\D^2 \rho)\xi,\label{tmkw3}
\end{equation}
where $\xi\coloneqq(2V-1)\rho^{-\frac{1}{2}}$,\ $\eta\coloneqq 2(2V+1)\rho^{-\frac{3}{2}}$ and $\tilde k_{ij}$ is the 
traceless part of $k_{ij}$. 
By the volume integration of these equations over $\Omega$ and the use of Stokes' theorem, 
one gets one equality and two inequalities. 
These inequalities are consistent with each other only when equalities hold in both.  
It gives the result that  $\Omega$ is spherically symmetric.
Finally, Ref. \cite{Xanthopoulos1991} 
shows that $\Omega$ is unique to be the BBMB solution. 

Compared to Eq. (\ref{tmkw1}), the derivation of Eqs. (\ref{tmkw2}) and (\ref{tmkw3})
is far from trivial. In the following, we will discuss the systematic way to derive them,
by applying the argument of Ref. \cite{Nozawa2018}. 

%
%======================================%
%<<<<<<<<<<<< SECTION III  >>>>>>>>>>>>%
%======================================%
\section{Generalization of the divergence equations and uniqueness}

In this section, we develop the systematic derivation of the divergence equations following Ref. \cite{Nozawa2018}. The obtained divergence equation allows us to 
show that the uniqueness of the photon surface of the BBMB solution. 

First, we wish to find a current $J^i$ satisfying the following equation.
\begin{equation}
D_i J^i =\left( \ \rm{terms \ a \ definite \ sign} \ \right).\label{divJ}
\end{equation}
The right-hand side is required to have a definite sign and to consist of a sum of tensors 
which vanish if and only if the spacetime is the BBMB solution. The candidate for such tensors is 
\begin{equation}
H_{ij}=D_i D_j v+\frac{3V}{1-V}D_i v D_j v-\frac{V}{\rho^2 (1-V)}g_{ij}\label{Hij}.
\end{equation}
Note that this is symmetric and traceless tensor. 
A simple calculation shows that $H_{ij}$ vanishes for the BBMB solution. 
The expression for $H_{ij}$ in terms of geometric quantities on $S_v$ is useful for later discussions.
\begin{eqnarray}
H_{ij}=\frac{1}{\rho}\Tilde{k}_{ij}-\frac{2}{\rho^2}n_{(i}\D _{j)}\rho
+\frac{1}{2\rho}(h_{ij}-2n_i n_j)\left(k-\frac{2V}{\rho(1-V)}\right)\label{Hij2},
\end{eqnarray}
where we used the Einstein equation. 
Using $H_{ij}$, we can also construct a vector $H_i$ which vanishes for the BBMB solution as  
\begin{eqnarray}
H_i=-\rho^2 H_{ij}D^jv
 \label{Hi}. 
\end{eqnarray}

Here we suppose that $J_i$ has the following form
\begin{equation}
J_i=f_1(v)g_1(\rho)D_i \rho+f_2(v)g_2(\rho)D_i v\label{Ji} .
\end{equation}
The divergence of Eq. (\ref{Ji}) is written by
\begin{eqnarray}
D_i J^i
& = & (f'_1 g_1 +f_2 g'_2)D_i v D^i \rho +f_1 g'_1 (D\rho)^2+f'_2 g_2 (Dv)^2+f_1 g_1 D^2 \rho \nonumber\\
& = &  -\rho^3f_1 g_1\left[|H_{ij}|^2 -\left(\frac{g'_1}{\rho g_1} + \frac{3}{\rho^2}\right)|H_i|^2\right]  
+ \rho f_1 g_1 D^i v H_i S_1   + \frac{1}{\rho^2}f_1 g_2 S_2,\label{divJ2}
\end{eqnarray}
where
\begin{equation}
S_1 \coloneqq \frac{f'_1}{f_1} + \frac{f_2}{f_1}\frac{g'_2}{g_1} + \frac{(4V-1)(3V-1)}{(2V-1)(1-V)}+
\frac{4\rho V}{1-V}\frac{g'_1}{g_1}\label{S1}
\end{equation}
and
\begin{equation}
S_2 \coloneqq \frac{2\rho V}{1-V} \frac{g_1}{g_2}S_1-\frac{4\rho V^2}{(1-V)^2}\frac{g_1}{g_2}\left[\rho \frac{g'_1}{g_1}+
\frac{8V^2-7V+2}{2V(2V-1)}\right]+\frac{f'_2}{f_1}.\label{S2}
\end{equation}
The prime denotes the differentiation with respect to each argument of the functions. 
In the second equality of Eq.(\ref{divJ2}), 
we have used \footnote{With aid of Eq. (\ref{einsteinij2}), the direct calculation from the definition of $\rho$ gives  this.}
\begin{equation}
D^2\rho=-\rho^3 |D_i D_j v|^2+\frac{3}{\rho}(D\rho)^2+\frac{1}{2V-1}\left(D_i \rho D^i v-\frac{4V}{\rho} \right).\label{D2rho}
\end{equation}
To control the sign of the right-hand side of Eq. (\ref{divJ2}), we require $S_1 = S_2 = 0.$  Following Ref. \cite{Nozawa2018}, 
to have decoupled equations, we suppose that $g_1$ and $g_2$ have the following form 
\begin{equation}
g_1=-c \rho^{-(c+1)},\ \ \ \ \ g_2 = \rho^{-c},\label{g-ansatz}
\end{equation}
where $c$ is an integration constant. Then, we have the two ordinary differential equations for $f_1$ and $f_2$ as 
\begin{equation}
f_2+f'_1 +\left[\frac{(4V-1)(3V-1)}{(2V-1)(1-V)}-\frac{4V(1+c)}{1-V}\right]f_1=0\label{odef1}
\end{equation}
and
\begin{equation}
f'_2+\frac{4cV^2}{(1-V)^2}\left[\frac{8V^2-7V+2}{2V(2V-1)}-(c+1)\right]f_1=0.\label{odef2}
\end{equation}
The solutions are given by 
\begin{equation}
f_1 = \frac{1}{4}(2V - 1)^{-1}(1 - V)^{1 - 2c}(a + b(2V - 1)^2)\label{solf1}
\end{equation}
and
\begin{equation}
f_2 = \frac{1}{4}(2V-1)^{-1}(1-V)^{-2c}\left[(a+b)(2cV-2V+1)-8bcV^2(1-V)\right], \label{solf2}
\end{equation}
where $a$ and $b$ are integration constants. Using the fact 
\begin{equation}
\frac{1}{2}\left|2\rho^2 H_{i[j} D_{k]}v - g_{i[j}H_{k]}\right|^2 = \rho^2|H_{ij}|^2 - \frac{3}{2}|H_i|^2,\label{arrangeHij}
\end{equation}
the divergence equation is rearranged as 
\begin{eqnarray}
D_i J^i  = \frac{c f_1}{2\rho^c}\left[\left|2\rho^2 H_{i[j} D_{k]}v - g_{i[j}H_{k]}\right|^2
+(2c-1)|H_{i}|^2\right].\label{divJ3}
\end{eqnarray}
To fix the sign of the right-hand side in Eq. (\ref{divJ3}), we require  
\begin{equation}
f_1 \geq 0,\ \ \ \ \ \ \ c \geq \frac{1}{2}.\label{cond-fc}
\end{equation}
With $\frac{1}{2} \leq V < 1$, it is easy to see that the former is guaranteed by
\begin{equation}
a \geq 0,\ \ \ \ \ a + b \geq 0.\label{cond-ab}
\end{equation}

Now, let us integrate Eq. (\ref{divJ3}) over $\Omega$. Using Stokes' theorem, we have  
\begin{equation}
\int_{\Omega} D_i J^i  d\Sigma = \int_{S_{\infty}}J_i n^i dS - \int_{S_p}J_i n^i dS \geq 0.\label{int-divJ}
\end{equation}
Using the asymptotic behaviors near the spatial infinity, $\rho\simeq |\partial_r V|^{-1}\simeq r^2/m$, $k\simeq 2/r$, 
the first terms of the right-hand side is estimated as 
\begin{equation}
\int_{S_{\infty}}J_i n^i dS=-\pi (a+b)m^{1-c}.
\end{equation}
For the second term, we have to carefully estimate it. Firstly, we have 
\begin{equation}
\int_{S_p}J_i n^i dS  =  -\frac{ac}{4}\Bigl(\frac{1}{2} \Bigr)^{1-2c}\frac{1}{\rho_p^{c+1}}\int_{S_p}\frac{k \rho_p-2}{2V-1}dS
-\frac{1}{4}\Bigl(\frac{1}{2} \Bigr)^{-2c}(a+b-2ac)\frac{1}{\rho_p^{c+1}}A_p, \label{preEstimateAtSp}
\end{equation}
where $A_p$ is the area of the surface $S_p$. Here note that the Gauss equation with the Einstein equation gives   
\begin{equation}
{}^{(2)}R= \frac{2}{\rho^2}+k^2-k_{ij}k^{ij}+\frac{2(\rho k -4V)}{(2V-1)\rho^2},\label{Geq}
\end{equation}
and then 
\begin{equation}
\lim_{V\to 1/2}{}^{(2)}R = \lim_{V\to 1/2}\frac{2(\rho k -2)}{(2V-1)\rho^2}\label{limtGeq}
\end{equation}
holds, where we used Eq. (\ref{BCatSp}). Using this and the Gauss-Bonnet theorem for the first term 
in the right-hand side of Eq. (\ref{preEstimateAtSp}), we arrive at
\begin{equation}
\int_{S_p}J_i n^i dS  =  -\frac{\pi ac}{2}\Bigl(\frac{1}{2} \Bigr)^{1-2c}\frac{1}{\rho_p^{c-1}} \chi
-\frac{1}{4}\Bigl(\frac{1}{2} \Bigr)^{-2c}(a+b-2ac)\frac{1}{\rho_p^{c+1}}A_p, \label{estimateAtSp}
\end{equation}
where $\chi$ is the Euler characteristic. As a consequence, Eq. (\ref{int-divJ}) implies  
\begin{align}
\label{}
(a+b)\left(A_p-\pi \rho_p^2 \left(\frac{4m}{\rho_p}\right)^{1-c}\right)
+ac \Bigl(\pi \rho_p^2 \chi-2 A_p \Bigr)\ge 0 \,. 
\end{align}
Under the parameter range (\ref{cond-fc}) and (\ref{cond-ab}), 
we get a pair of inequalities
\begin{align}
\label{}
\pi \rho_p^2 \left(\frac{4m}{\rho_p}\right) ^{1-c} \le A_p \le \frac 12 \pi \rho_p^2 \chi \,. 
\end{align}
Setting $c=1$ gives $\chi \ge 2$, meaning that only the allowed topology of $S_p$ is 
spherical ($\chi=2$). Setting $\chi=2$ implies that 
the equality holds, and it occurs if and only if $H_{ij}$ vanishes. This is the case that the spacetime is 
spherically symmetric. According to Ref. \cite{Xanthopoulos1991}, the spacetime is unique to be 
the BBMB solution. 

%\begin{equation}
%-\pi(a+b)\Bigl(\frac{4m}{\rho_p} \Bigr)^{1-c}\rho_p^2+ac\pi\rho_p^2\chi+(a+b-2ac)A_p \geq 0. \label{ineq}
%\end{equation}
%For $a=1$ and $b=-1$, the above inequality is rearranged as 
%\begin{equation}
%A_p \leq \frac{\pi \chi}{2}\rho_p^2.
%\end{equation}
%Then, we see $\chi>0$ and the topology of $S_v$ is isomorphic to 2-sphere ($\chi=2$). 
%Thus, we have $A_p \leq \pi \rho_p^2$. 
%For the case with  $a=0,$ and $c=1$, Eq. (\ref{ineq}) gives $A_p \geq \pi \rho_p^2$. 

Before closing this section, we comment on the relation to the divergence identities in Ref. \cite{Tomikawa2017a}.
For $b=0$ and $c=1/2$, Eq. (\ref{divJ3}) coincides with Eq. (\ref{tmkw2}), and for $a=0$ and $c=1/2$, 
Eq. (\ref{tmkw3}). See Appendix for the details. 

%
%======================================%
%<<<<<<<<<<<< SECTION VI  >>>>>>>>>>>>>%
%======================================%
\section{Summary}

In this paper, we reexmined the Israel-type proof for the uniqueness of the photon surface in 
the Einstein-conformal scalar system. Following Ref. \cite{Nozawa2018}, we have derived a new 
divergence identity with three parameters and given a new proof for the uniqueness. In Ref. \cite{Nozawa2018}, 
vacuum Einstein, Einstein-Maxwell and Einstein-Maxwell-dilaton systems have been addressed. 
Therefore, the current study indicates the powerfulness of the systematic procedure presented there. 
The deep physical/mathematical reason is expected to be hidden behind the presence of such a procedure. 

\ack

We would like to thank Masato Nozawa for valuable comments to the draft.
K. I. and T. S. are supported by Grant-Aid for Scientific Research from 
Ministry of Education, Science, Sports and Culture of Japan (No. JP17H01091). K.~I. is also 
supported by JSPS Grants-in-Aid for Scientific Research (B) (JP20H01902). T. S. is also supported 
by JSPS Grants-in-Aid for Scientific Research (C) (JP21K03551). 

%
%======================================%
%<<<<<<<<<<<<<< appendix  >>>>>>>>>>>>>%
%======================================%

\appendix

\section{Relation of Eq. (\ref{divJ3}) to Eqs. (\ref{tmkw2}) and (\ref{tmkw3})}

Using Eq. (\ref{tmkw1}) or equivalent equality 
\begin{equation}
k \rho = n^i D_i \rho, \label{kr}
\end{equation}
we have 
\begin{equation}
J_i=\frac{-cf_1\rho k+f_2}{\rho^{c+1}}n_i-cf_1\frac{{\cal D}_i \rho}{\rho^{c+1}}.
\end{equation}
Using 
\begin{equation}
D^i \Biggl( \frac{{\cal D}_i\rho}{\rho^{c+1}}\Biggr)=\frac{{\cal D}^2\rho}{\rho^{c+1}}-c \frac{({\cal D}\rho)^2}{\rho^{c+2}},
\end{equation}
the left-hand side of Eq. (\ref{divJ3}) becomes
\begin{equation}
D^i J_i= D^i \Biggl( \frac{-cf_1\rho k+f_2}{\rho^{c+1}}n_i \Biggr)
-cf_1\frac{{\cal D}^2\rho}{\rho^{c+1}}+c^2f_1\frac{({\cal D}\rho)^2}{\rho^{c+2}}.
\end{equation}
The right-hand side of Eq. (\ref{divJ3}) is expressed as  
\begin{equation}
c \frac{f_1}{\rho^c}\Biggl[ \tilde k_{ij} \tilde k^{ij}+\frac{c}{\rho^2}({\cal D}\rho)^2+\frac{2c-1}{2}
\Bigl(k-\frac{2V}{\rho(1-V)} \Bigr)^2  \Biggr].
\end{equation}
Thus, we have 
\begin{equation}
D^i \Biggl( \frac{-cf_1\rho k+f_2}{\rho^{c+1}}n_i \Biggr)=c \frac{f_1}{\rho^c}\Biggl[ \tilde k_{ij} \tilde k^{ij}+\frac{2c-1}{2}\Biggl(k-\frac{2V}{\rho(1-V)} \Biggr)^2  \Biggr]
+cf_1\frac{{\cal D}^2\rho}{\rho^{c+1}}.\label{divJ4}
\end{equation}

Now we consider the $c=1/2$ case and then $f_1=(1/4)(2V-1)^{-1}[a+b(2V-1)^2]$ and $f_2=(1/4)(2V-1)^{-1}[a+b(1-2V)(1+2V)]$. 
Setting $b=0$($a=0$), we can see that Eq. (\ref{divJ4}) becomes Eq. (\ref{tmkw2}) (Eq. (\ref{tmkw3})).

\end{document}